\documentclass[aps,prl,preprint,amssymb,showpacs,superscriptaddress,]{revtex4}
\usepackage{graphicx}% Include figure files
\usepackage{dcolumn}% Align table columns on decimal point
\usepackage{bm}% bold math
\usepackage{color}

\begin{document}

\title{Spin decoherence and electron spin bath noise of a nitrogen-vacancy center in diamond}

\author{Zhi-Hui Wang}
\affiliation{Department of Chemistry, University of Southern California, Los Angeles CA 90089, USA}
\affiliation{Center for Quantum Information Science and Technology, University of Southern California, Los Angeles, California 90089, USA}
\author{Susumu Takahashi}
\email{susumu.takahashi@usc.edu}
\affiliation{Department of Chemistry, University of Southern California, Los Angeles CA 90089, USA}
\affiliation{Department of Physics, University of Southern California, Los Angeles CA 90089, USA}
\affiliation{Center for Quantum Information Science and Technology, University of Southern California, Los Angeles, California 90089, USA}

\date{\today}
\begin{abstract}
We theoretically investigate spin decoherence of a single nitrogen-vacancy (NV) center in diamond. Using the spin coherent state P-representation method, coherence evolution of the NV center surrounded by nitrogen electron spins (N) is simulated. We find that spin decoherence time as well as free-induction decay of the NV center depend on the spatial configuration of N spins. Both the spin decoherence rate (1/$T_2$) and dephasing rate (1/$T_2^*)$ of the NV center increase linearly with the concentration of the N spins.
Using the P-representation method, we also demonstrate extracting noise spectrum of the N spin bath, which will provide promising pathways for designing an optimum pulse sequence to suppress the decoherence in diamond.
\end{abstract}

\maketitle

%\section{Introduction}
%%%%%%%%%%%%%
%%% NV introduction
%%%%%%%%%%%%%
Nitrogen-vacancy (NV) impurity centers in diamond have been investigated extensively for fundamental research~\cite{gruber97, Jelezko02}
and potential applications of quantum information processing devices~\cite{JelezkoGate04, Jiang09, Neumann10},
and a high-precision room-temperature magnetic sensor~\cite{Maze08, Balasubramanian08}.
Coherent properties of NV centers in diamond play a crucial rule in the applications. Spin decoherence is due to couplings to noisy environments.
For a NV center in diamond, major noise sources are paramagnetic impurities and nuclear spins.
Diamond crystals containing $>$ 10 ppm of substitutional single-nitrogen (N) impurities are called type-Ib diamonds and
the spin decoherence is caused by fluctuating N spin baths~\cite{kennedy03, Takahashi08, hanson08}.
On the other hand, in diamond containing much less N spins, the spin decoherence time $T_2$ is much longer and the decoherence is often
limited by couplings to $^{13}$C nuclear spins ($\sim$1.1 $\%$ natural abundance)~\cite{childress06, Maze08b, Balasubramanian08}.
%In addition, the decoherence has been suppressed using dynamical decoupling (DD) techniques in which the coupling to the environmental noise is reduced by optimizing a pulse sequence for pulsed electron spin resonance (ESR) spectroscopy.

Understanding dynamics of an electron spin system and surrounding electron and nuclear spin baths has been a long-standing problem for theoretical investigation on electron spin resonance (ESR) in solids. The lineshape and width of continuous ESR spectrum as well as pulsed ESR signals including free induction decay (FID) and spin echo (SE) decay were successfully described by approximating spin baths by stochastic noise fields~\cite{KlauderAnderson, Chiba72, Zhidomirov69, deSousa03}.
In order to realize spin-based quantum bits (qubits) in solid-state systems~\cite{Kane98, Loss98}, investigation of spin baths is becoming more critical to understand decoherence. Methods like dynamical decoupling~\cite{viola98, khodjasteh05, uhrig07} has been studied to suppress the decoherence in knowledge of noise spectrum of surrounding spin baths~\cite{medford12, Cywinski08PRB}.
In semiconductor quantum dots~\cite{hanson07}, silicon~\cite{MortonReview11}, and diamond with very low concentration of N spins~\cite{childress06, Maze08b, RBLiu12}, nuclear spins are the major source of the decoherence in the system. The dipolar coupling between nuclear bath spins is much weaker than the hyperfine (HF) coupling
between an electron (central) spin qubit and bath nuclear spins (the system-bath coupling).
In such cases, microscopic treatments of nuclear spin baths have been thoroughly studied using the cluster expansion method~\cite{Witzel05, Witzel06, Saikin07, RBLiu06, RBLiu07}
in which the intra-bath coupling is treated as a perturbation to the system-bath coupling.
On the other hand, for electron spin baths where the strength of the intra-bath coupling is comparable to that of the system-bath coupling (the strong intra-bath coupling regime),
it is often challenging to implement the cluster expansion method~\cite{Witzel10, Witzel12}.
In type-Ib diamond, the dipolar coupling strength between the NV center and the N spins is of the same order of that between N spins.
Although it has been shown that experimentally observed FID and SE signals of a single NV center have been well described by a classical stochastic noise, the Ornstein-Uhlenbeck (O-U) process~\cite{hanson08, deLange10},
spin dynamics of a NV center and surrounding N spins has not been fully understood yet.

Variations of the P-representations~\cite{glauber63, sudarshan63, drummond80} have been successfully used to formulate many-body problems in quantum optics, {\it e.g.} squeezing in optical solitons~\cite{carter87}.
The spin coherent state P-representation, as a variation of the time-dependent mean-field method, has
been proposed to apply to spin-based qubit systems targeting at decoherence of the central spin~\cite{PrepPRL06, Zhang07JOP}.
The wavefunction of the whole system is represented in the basis of direct product of wavefunctions for individual spins.
Equations of motion for all spins are specially tailored to achieve a close approximation to quantum dynamics of the central spin.
Simulation of FID and SE for a nuclear spin system has been demonstrated using the spin coherent P-representation method~\cite{Zhang07JOP}.

In this paper, we theoretically study spin decoherence of a single NV center in diamond with the N electron spin bath.
We employ the P-representation method to simulate spin dynamics of the NV center and surrounding N spins.
Simulated SE decays agree with the decays due to a classical noise field described by the O-U process.
Our simulation shows that the decay rate of both FID ($1/T^*_2$) and SE ($1/T_2$)
depend linearly on the concentration of N spins in the range from 1 to 100 ppm.
Noise spectrum of N spins is also extracted using the P-representation method.
The simulated noise spectrum is in good agreement with the noise spectrum of the O-U process in the range of high frequencies
with which the spectrum of the SE sequence overlaps significantly.

%\section{Spin Hamiltonian}
We consider a NV center in diamond ($S$ = 1) under a static magnetic field $B_0$ applied along the N-V axis (denoted as the $z-$axis).
The Hamiltonian of the NV center system is
\begin{eqnarray}
H_S=D(S_0^z)^2+\gamma_0 B_0 S^z_0
\end{eqnarray}
where $D$ = 2.87 GHz is the zero-field splitting due to the axial crystal field, $\gamma_0$ is the gyromagnetic ratio for the NV center, and $S_0$ is the electron spin of the NV center. The second term corresponds to the Zeeman energy of $S_0$.
The HF coupling between $S_0$ and the nitrogen nuclear spin in the NV center is not considered.
The degeneracy between $m_z=+1$ and $m_z=-1$ states is lifted by the external magnetic field $B_0$.
We consider only the $m_z=-1$ and $m_z=0$ transition to treat the NV center as a two-level system, $s_0$.
The Hamiltonian for an individual N spin ($S$ = 1/2) is
\begin{eqnarray}
\label{eq:Hami}
H_k=\gamma B_0 S^z_k+ A_1S^z_kI^z_k
\end{eqnarray}
where $\gamma$ is the gyromagnetic ratio for the N spin and the HF coupling constant is $A_1 = 114$ MHz for N spins delocalized along the $[111]$ axis and 86 MHz along the $[11\bar{1}]$, $[1\bar{1}1]$ and $[\bar{1}11]$ axes~\cite{Loubser78}.
We here consider cases where $B_0$ is not equal to 514 G, therefore there is a large mismatch in the transition energy of NV and N spins which suppresses the flip-flop process between NV and N spins greatly.
In the rotating frame with the precession frequency of the N spins, $\gamma B_0 S_k$, and with the NV center, $(D-\gamma_0 B_0) s_0$, the Hamiltonian for the dipolar interaction between the NV and the N spins (the system-bath coupling) is given by~\cite{hanson08},
\begin{eqnarray}
H_{SB}=(s_0^z-1/2)\sum_k A_k S^z_k
\end{eqnarray}
where $A_k=[1-3(n^z_k)^2]a_k$ and $a_k=\hbar\gamma_0\gamma/r_k^3$ is the coupling constant.
The Hamiltonian for the dipolar coupling between N bath spins $S_j$ and $S_k$ is given by
\begin{eqnarray}
H_{B}=\sum_{j,k} c_{j,k}[1-3(n^z_{j,k})^2] [S^z_jS^z_k-\frac{1}{4}(S^+_jS^-_k+S^-_jS^+_k)]
\end{eqnarray}
where $c_{j,k}=\hbar\gamma^2/r_{j,k}^3$ and only secular terms are considered.

\begin{figure}
\includegraphics[width=100 mm]{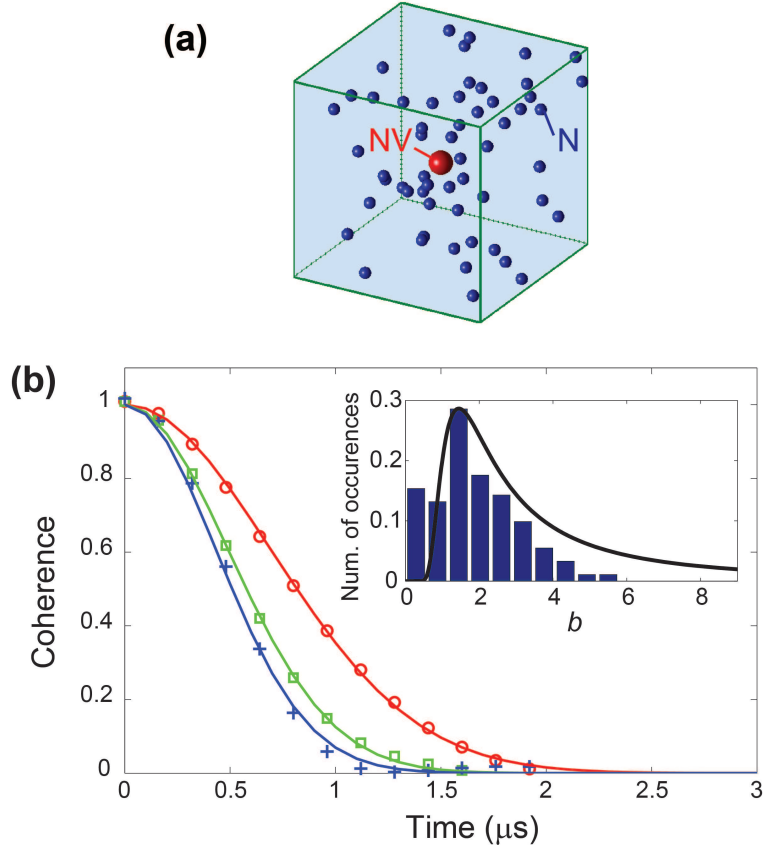}
\caption{
\label{Fig1}
(a) A Schematic of a spatial configuration for a NV and N spins.
The cube consists of the diamond lattice. A NV center (red large sphere) is located at $(0,0,0)$
and N spins (blue small spheres) are located randomly on diamond lattice sites.
(b) Simulated FID signals of a NV center with three different N spin bath configurations.
The concentration of N spins is $f = 10~$ppm for all cases.
Blue crosses, green squares and red circles are simulated results, and lines are fit to $\exp[-(t/T^*_2)^2]$.
Obtained $T^*_2=0.97$, $0.69$, and $0.61~\mu s$ from the fit are in good agreement with
calculated values directly from the N spin bath configuration where
$T^*_2=\sqrt 2/b$ are $0.95$, $0.67$, and $0.62$ $\mu s$ respectively.
The inset shows a histogram of $b$ for the 90 instances with $f=10$~ppm.
Black solid line represents the expected probability distribution of $b$ (see the supplemental material for details).
}
\end{figure}
%\section{P-representation Method}
Fig.~\ref{Fig1} shows a spatial configuration of a NV center ($S_0$) and N spin ($S_k$) bath used in our simulation.
The NV center spin is located at the center of a cube where the cube consists of a tetrahedral diamond unit cell with $a = 3.567\AA$ of the lattice constant. The number of the unit cells in the cube is $N_a^3$ ($N_a$ is the number of unit cell along the $x$,$y$ and $z$-axes).
N spins are randomly distributed on the lattice sites. $N_a$ and the number of N spins (typically 80 $\sim$ 100 N spins are used) are adjusted according to the concentration of N spins ($f$). In the simulation, M initial vectors are sampled to represent unpolarized N spin bath corresponding to a high temperature limit of the bath, $T\gg \hbar\gamma B_0/k_B$, $\big({\bf R}_0^{(m)},{\bf R}_1^{(m)},\cdots,{\bf R}_J^{(m)}$) and
$m=1,\ldots M$, are prepared where $J$ is the number of N spins and ${\bf R}_k=(\theta_k,\phi_k)$ is a classical vector on the Bloch sphere of spin $k$ (the $k$=0 spin is a NV center). Time evolution of the vector, ${\bf R}_k(t)$, is calculated according to a set of equations of motion
$\dot {\bf R}_k={\bf B}_{k}\times {\bf R}_k$  where ${\bf B}_{k}$ is a local magnetic field for spin $k$ induced by other spins in the system.
Coherence at time $t$ is given by calculating $\langle s^x_0(t) \rangle = \frac{1}{M}\sum_{m=1}^M R^{x,(m)}(t)$ (see the supplemental material for details).

Simulated coherence are shown in Fig.~\ref{Fig1}(b).
The coherence decays as a function of time, and the decay corresponds to FID signals of a NV center.
As shown in Fig.~\ref{Fig1}(b), simulated FIDs are fit well by the fundamental Gaussian function.
The Gaussian shape of the FID agrees with that of FID signals experimentally observed from a single NV center in type-Ib diamond~\cite{hanson08}.
The root-mean-square of the spin-bath coupling $b=\frac{1}{2}\sqrt{\sum_j A^2_j}$ quantifies the FID for the NV spin, {\it i.e.} $T^*_2=\sqrt 2/b$.
Thus the FID time depends on the configuration of the local bath spins around the NV center.
Fig.~\ref{Fig1}(b) shows the simulated FID signals for the single NV center with different N bath configurations for $f=10~$ppm of the N spin concentration.
We found that the values of $b$ obtained from $T^*_2$ agree with $b$ directly calculated from the configuration of the N bath spins. Inset of Fig.~\ref{Fig1}(b) shows a histogram of $b$. The distribution of $b$ agrees with the theoretically expected distribution~\cite{Dobrovitski08}.
To represent a typical configuration of the N spin bath, simulated results with any pairs of spins coupled much stronger (50 times) than typical coupling strength are excluded. A deviation at high $b$ values is due to this exclusion (see the inset of Fig.~\ref{Fig1}(b)).
We obtained $T^*_2$ = $0.02\sim 1.73~\mu s$ with 0.53~$\mu s$ of the the mean value for $f=10~$ppm of the N spin concentration.

\begin{figure}
\includegraphics[width=120 mm]{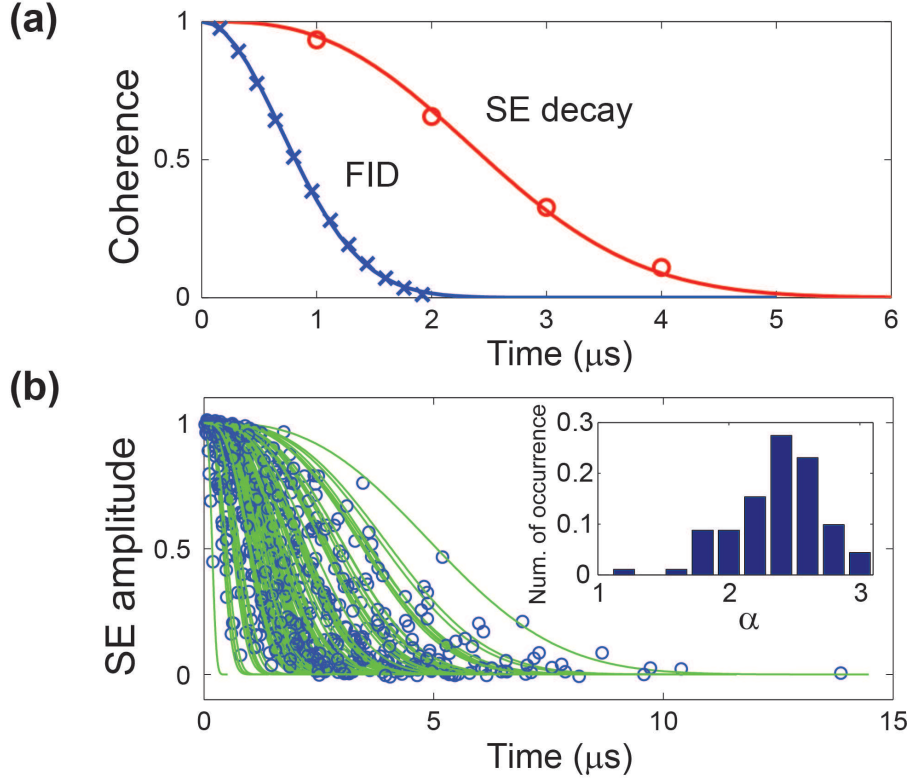}
\caption{
\label{Fig2}
(a) Simulated SE and FID signals with $f=10~$ppm.
Circles and crosses are simulation results, and a red solid line shows a fit to Eq.~(\ref{eq:T2KA}).
With $b=1.44~\mu s^{-1}$ from the FID data, we obtained $\tau_C$ = $2.78~\mu s$.
(b) Simulated SE signals for 90 instances of the N spin bath configuration with $f=10~$ppm.
Blue circles are simulation results, and green lines are
fits to Eq.~(\ref{eq:T2KA}).
The inset shows a histogram of $\alpha$ for the 90 instances when
SE signals are fitted by $\exp[-(t/T_2)^\alpha]$.
}
\end{figure}

As shown in Fig.~\ref{Fig2}(a), we simulated the time evolution of SE.
The rephasing $\pi-$pulse is assumed to be perfect and instantaneous in the simulation.
SE decay in electron spin baths has been described by treating
the bath to be a classical noise field where
the noise field $B(t)$ was modeled by the O-U process
with the correlation function $C(t) = \langle B(0)B(t)\rangle = b^2 \exp(-|t|/\tau_C)$.
$\tau_C$ is the correlation time of the bath, which measures the rate of the flip-flop process between the bath spins.
%{\color{blue}The O-U process is a Gaussian, stationary, and Markovian process. THIS IS NOT CLEAR.}.
%{\color{green}COMMENTS: How about we remove this line since it is a general statement about O-U process which is a well-defined and well-understood stochastic process and is not our main content.}
The corresponding noise spectrum is Lorentzian with power $b^2$ and the half-width at half-maximum (HWHM) $1/\tau_C$.
SE decay subject to the noise due to the O-U process
is given by~\cite{KlauderAnderson},
\begin{eqnarray}
\label{eq:T2KA}
E(t)&=&\exp[-(b\tau_C)^2[t/\tau_C-3-e^{-t/\tau_C}+4e^{-t/(2\tau_C)}]].
\end{eqnarray}
In the quasi-static limit ($b\tau_c \gg 1$) indicating slow bath dynamics, $E(t)=e^{-b^2t^3/(12\tau_C)}\sim\exp[-(t/T_2)^3]$.
On the other hand, in the motional-narrowing limit ($b\tau_c \ll 1$),
$E(t)=e^{-t/\tau_C}\sim\exp(-t/T_2)$.
As shown in Fig.~\ref{Fig2}(a), we found good agreement between our simulation results and Eq.~(\ref{eq:T2KA}).
Using the value of $b$ determined by the bath configuration and confirmed by the FID,
we determined $\tau_C$.
Fig.~\ref{Fig2}(b) shows 90 simulated SE decays with $f$=10 ppm.
For many cases, we found $b\tau_C>1$, and SE decays are well described by $\exp[-(t/T_2)^\alpha]$
where the exponent $\alpha$ is typically between $2\sim 3$ as shown in inset of Fig.~\ref{Fig2}(b).
This is in consistence with previous experimental results ($\alpha \sim 3$)~\cite{deLange10, ZHNVDD}.
We also found a few cases in the motional-narrowing regime ($b\tau_C<1$),
in which the SE decay shapes are close to a single exponential function.

\begin{figure}
\includegraphics[width=120 mm]{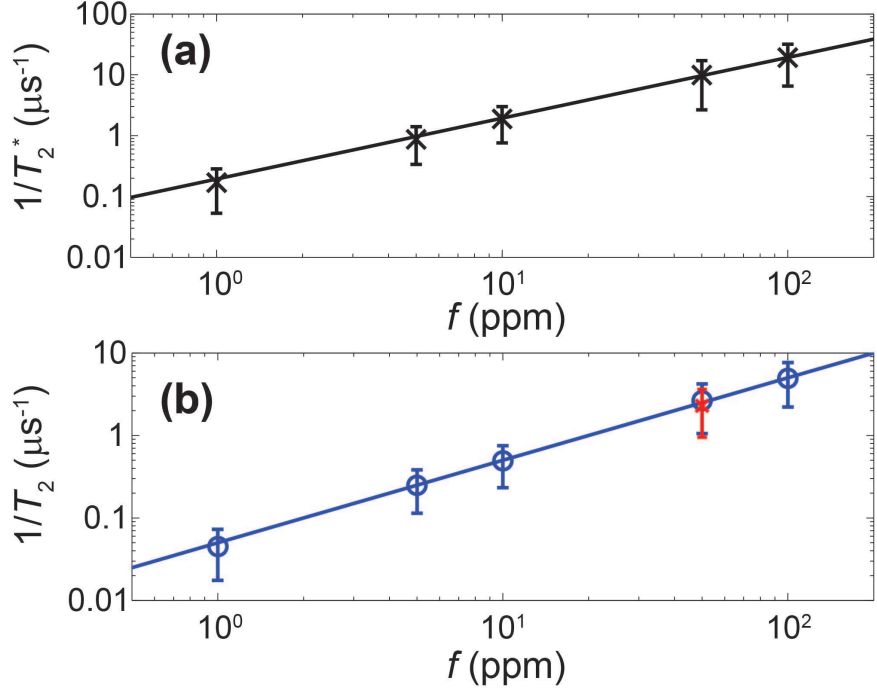}
\caption{
\label{Fig3}
(a) $1/T^*_2$ as a function of the N spin concentration.
Each cross is the mean value and the error bar is the standard deviation of all instances.
Black solid line is a fit to a linear function.
(b) $1/T_2$ as a function of the N spin bath concentration. Each blue circle is the mean value and the error bar is the standard deviation of all instances. Blue solid line is a fit to a linear function.
The red cross and its error bar are the results with the HF interaction between N electron and $^{14}$N nuclear spins taken into account. Difference in simulated $1/T_2$ between with and without the HF coupling is smaller than the deviation (see the supplemental material for details).
}
\end{figure}

%\subsubsection{$T_2$ and $T^*_2$ dependence on the concentration of P1 center}
We examined the concentration dependence of the FID and SE decay in the range from $1$ to $100$ ppm. In Fig.~\ref{Fig3}(a) and (b), $1/T^*_2$ and $1/T_2$ are shown for the concentrations of $f = 1,~5,~10,~50,~100$ ppm. For each value of $f$, $80\sim 100$ random configurations of the bath are simulated and the mean of $1/T^*_2$ and $1/T_2$ with the standard deviation as an error bar are shown. We found a linear dependence on $f$ for both $1/T^*_2$ and $1/T_2$, namely $1/T^*_2 = 0.19f$ and $1/T_2 = 5\times10^{-2} f$ corresponding to $T_2 = 2.03$ and $0.2~\mu s$ for $f$ = 10 and 100 ppm respectively.
The linear dependence of the concentration has been experimentally seen in $1/T_2$ of ensemble N spins in diamond~\cite{Wyk97}.
In addition, a linear concentration dependence of $1/T_2$ has been reported in phosphorous donors in silicon~\cite{Witzel10}.
% and also has been studied theoretically for sparse paramagnetic system using methods of moment.~\cite{Abragam}
The linear dependence can be understood as stemming from the dipolar nature of both the spin-bath and the intra-bath couplings. Consider the case of $b\tau_C\gg 1$. The average spin-bath coupling $\overline A_j \propto 1/\bar r^3\propto f$, hence $\overline {b^2}\propto f^2$, where $\bar r$ is the average distance between neighboring spins.
Similarly for N bath spins, $\overline {\tau_C}$ is roughly proportional to average coupling, which is proportional to $1/f$. For $b\tau_C>1$, the expansion of Eq.~\ref{eq:T2KA} to the leading order in $t/\tau_C$ yields $\overline {1/T_2}\propto(\overline{b^2}/\overline{\tau_C})^{1/3}$, therefore $1/T_2$ is proportional to $f$.

\begin{figure}
\includegraphics[width=120 mm]{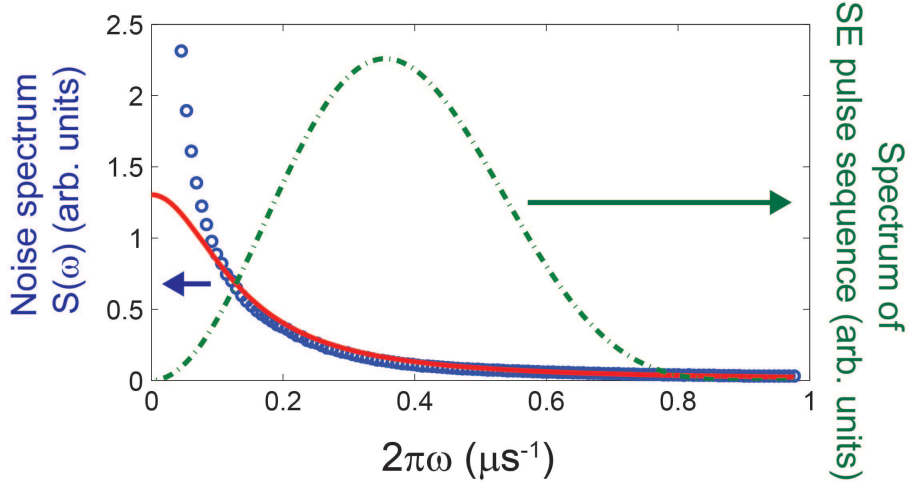}
\caption{
\label{fig:noiseSpec}
Noise spectrum of N spins with $f=10~$ppm.
Blue circles are the simulated noise spectrum.
Red solid line shows the noise spectrum of the O-U process expressed by the Lorentzian function with $1/\tau_C=0.134~\mu s^{-1}$ where $\tau_C$ is extracted from the SE signal.
Black dotted line shows the power spectrum of SE pulse sequence $|\sin^2(\omega t/4)/(\omega/4)|^2$ with $t = T_2 = 3.8\mu s$.
}
\end{figure}

%{\bf Noise spectrum}
Finally we simulated noise spectrum of the N spin bath, $S(\omega)$. At time $t$, the local magnetic field due to N spins at the NV spin is expressed by $B_{NV}(t) \sim \sum_{j} A_{0,j}\langle R^z_j(t)\rangle$ where the sum is taken over all N bath spins. Direct Fourier transform of $B_{NV}(t)$ renders the noise spectrum, $S(\omega)$. Fig.~\ref{fig:noiseSpec}
shows the simulated noise spectrum of N spins.
Coherence of the NV center decays due to couplings to environmental noise of N spins. The amount of the decay is determined by overlaps between the noise spectrum and the spectrum of a pulse sequence used in ESR measurements. In SE measurement, the power spectrum of the pulse sequence is given by $|\sin^2(\omega t/4)/(\omega/4)|^2$ where $t$ is the total evolution time~\cite{Cywinski08PRB}.
As shown in Fig.~\ref{fig:noiseSpec}, with $t=T_2$, significant overlap between the noise and the pulse-sequence spectrum happens at high frequencies where the noise spectrum agrees well with the spectrum of the O-U process. This supports the observation of good agreement between our simulated SE signals and the analytical solution with the O-U process (see Eq.~\ref{eq:T2KA}). On the other hand, in the present case, the dephasing time $T_2^*$ of the FID signals is determined by $e^{-t^2\int^{+\infty}_{-\infty}S(\omega)d\omega}$ where $T_2^*$ does not depend on details of the noise spectrum $S(\omega)$. We found excellent agreement in between $T_2^*$ obtained from the noise spectrum and $T_2^*$ extracted from simulated FID.

In summary, we investigated spin decoherence of
a single NV center coupling to N electron spins using the P-representation method.
The P-representation is a modified time-dependent mean-field technique that is suitable to simulate coherence evolution of a central spin surrounded by spin baths.
Simulated results for SE signals are in good agreement with analytical expressions based on the O-U process and with previous experimental results.
We found that the FID and SE decay rates, $1/T_2^*$ and $1/T_2$, depend linearly
on the concentration of N spins.
We also demonstrated simulating the noise spectrum of the electron spin bath using the P-representation method.
The P-representation method is suitable to study decoherence in other spin-based qubit systems even in the strong intra-bath coupling regime, {\it e.g.} phosphorus donors in isotropically pure silicon ($^{29}$Si $< 50 $ppm).
This capability will facilitate us to find optimum DD sequences to effectively suppress decoherence in spin-based qubit systems.

We thank to V. V. Dobrovitski and D. A. Lidar for useful discussions. This work was supported by the Searle scholars program (S.T.).

%=======================================
%\bibliography{PrepNV}

\end{document}

{\center{\large{\bf SUPPLEMENTAL MATERIAL}}}

\vspace{3mm}

\title{Spin decoherence and electron spin bath noise of a nitrogen-vacancy center in diamond}

\author{Z.-H. Wang and S. Takahashi}

\makeatletter

\renewcommand{\thefigure}{S\@arabic\c@figure}

%%%%%%%%%%%%%%%%%%%%%%%%%%%%%%%%%%%%%%%%%%%%%%%%%%%%%%%%%%%%%%%%%%%%%%%%

\maketitle

%%%%%%%%%%%%%%%%%%%%%%%%%%%%%%%%%%%%%%%%%%%%%%%%%%%%%%%%%%%%%%%%%%%%%%%%%
Here we provide details of the method to simulate spin dynamics of a single nitrogen-vacancy (NV) center and surrounding single-substitutional nitrogen (N) spins, and discussion of the hyperfine (HF) coupling between N electron and $^{14}$N nuclear spins and the spin bath configuration dependence of the root-mean-square of the spin-bath coupling ($b$).

\vspace{3mm}

{\bf (i) P-representation Method}

Free-induction decay (FID) and spin echo (SE) decay signals of the NV center, and noise spectrum of N spins in diamond are calculated using the P-representation method.
In our simulation, a single NV center is positioned at the center of a cube consisting of a diamond lattice, and N spins are randomly located on diamond lattice sites. The number of the unit cells in the cube and the number of N spins are adjusted according to the concentration of N spins ($f$).
We consider the magnetic dipole interaction between all pairs of spins in the system to simulate their spin dynamics. In order to represent typical cases, we excluded bath configurations in which a dipolar coupling strength between any pairs of spins is more than 50 times larger than the typical coupling amplitude, $g^2\mu_B^2f/a^3$, where $g=2$ is the electron $g$-factor, $\mu_B$ is the Bohr magneton, and $a=3.567\AA$ is the lattice constant of the diamond lattice.

Initially $M=40000$ sets of random vectors
$\big({\bf R}_0^{(m)},{\bf R}_1^{(m)},\cdots,{\bf R}_N^{(m)}$),
$m=1,\ldots M$ representing unpolarized initial states $\rho(0)=\frac{1}{2^{N+1}}|x\rangle\langle x|\otimes_{k=1}^N {\mathbf 1}_k$ are prepared~\cite{Zhang07JOP}
where ${\bf R}_k=(\theta_k,\phi_k)$ represents a classical vector of spin $k$ on the Bloch sphere.
Then we compute time evolution of the sample vectors according to a set of equations of motion,
$\dot {\bf R}_k={\bf B}_{k}\times {\bf R}_k$, where ${\bf B}_{k}$ is a local magnetic field for spin $k$ created by all other spins in the system~\cite{PrepPRL06}.
For the NV center spin ($k=0$), ${\bf B}_{k}=\sqrt{3}/2\sum_{k'\ne 0} A^z_{k,k'}R^z_{k'}\hat{\bf z}$ where
$A^z_{k,k'}$ is the $z$ component of the dipolar coupling to spin $k$. The x and y components are ignored because of a negligible flip-flop rate between NV and N spins.
For bath spins ($k=1,\ldots N$),
$B^{\beta}_{k}=\sqrt{3}/2\big(\sum_{k'\ne k} A^{\beta}_{k,k'}R^{\beta}_{k'}-\frac{1}{2}A^z_{k,0}\delta_{\beta,z}\big)$
where $\beta=x,y,z$
and the second term, $1/2A_{k,0}$, is resulted from the mapping of the NV center spin $S=1$ to the effective spin of $s_0=1/2$ (see Eq.~(3) in the main text).
Coherence at time $t$ is computed as $\langle s^x_0(t) \rangle = \frac{1}{M}\sum_{m=1}^M R^{x,(m)}(t)$

{\bf (ii) Effect of the HF interaction between N electron and $^{14}$N nuclear spins}

Most of N electron spins interact with $^{14}$N nuclear spin (I = 1, the natural abundance of $^{14}$N is 99.6$\%$) via the HF interaction. The HF interaction splits the energy levels of the N electron spin. Because of a large energy mismatch between N electron and nuclear spins, the flip-flop process between the electron and nuclear spin is largely suppressed, leaving the effective HF term in Hamiltonian of the system to be $A_kS^z_kI^z_k$.
The HF interaction provides a static detuning magnetic field for N electron spins. In the presence of the static field $B_0$ along $[111]$, the HF coupling is $114~$MHz for the N spin delocalized along the $[111]$ and $86~$MHz along the $[\bar{1}11]$, $[1\bar{1}1]$ and $[11\bar{1}]$, therefore the N spin state is split into five states with 1:3:4:3:1 population ratio.
While N electron spins in the same energy state can flip-flop due to the dipolar interaction, the flip-flop rate between electron spins with different states are negligible due to the large energy difference.
\begin{figure}
\includegraphics[width=7cm]{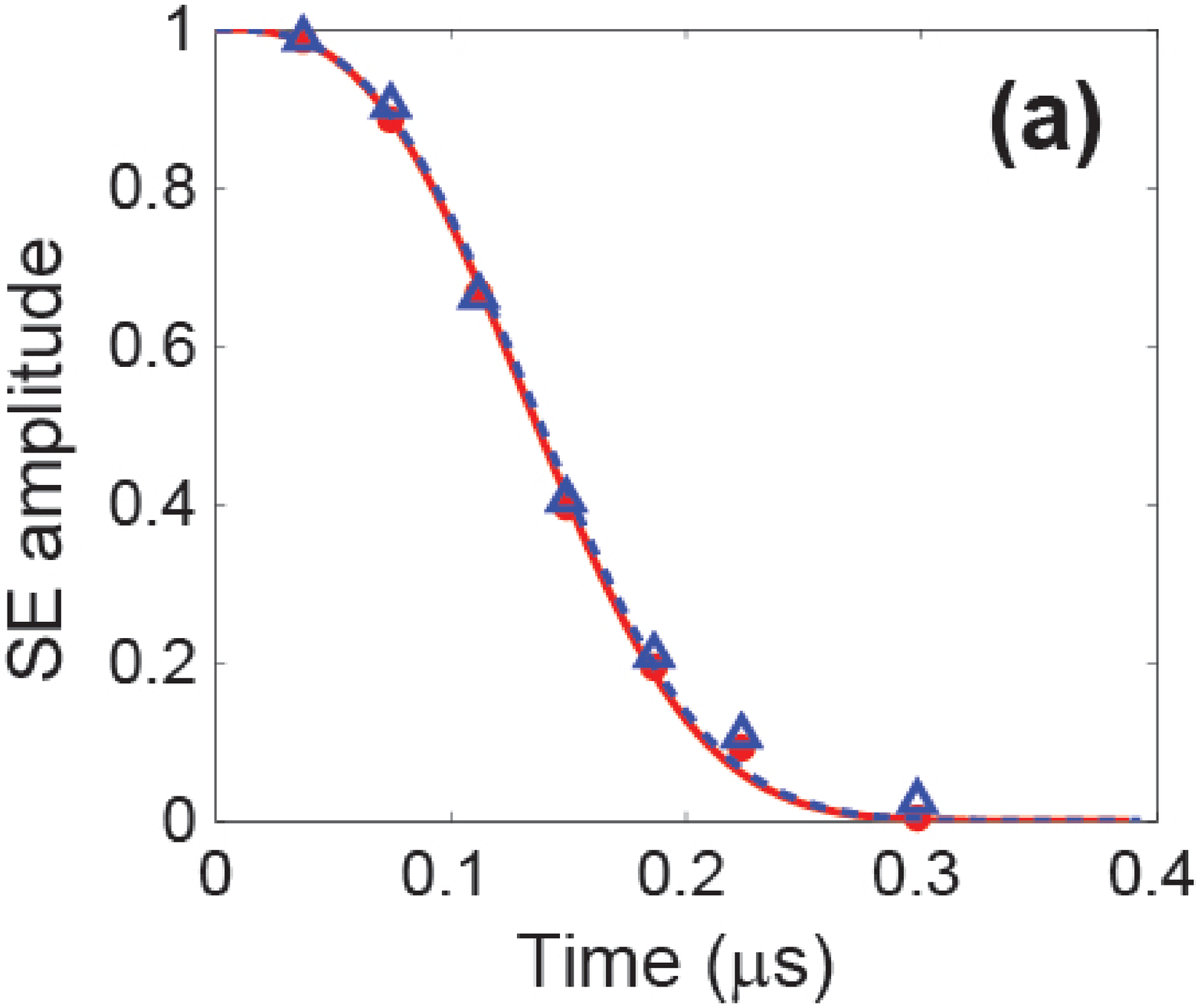}
\includegraphics[width=7cm]{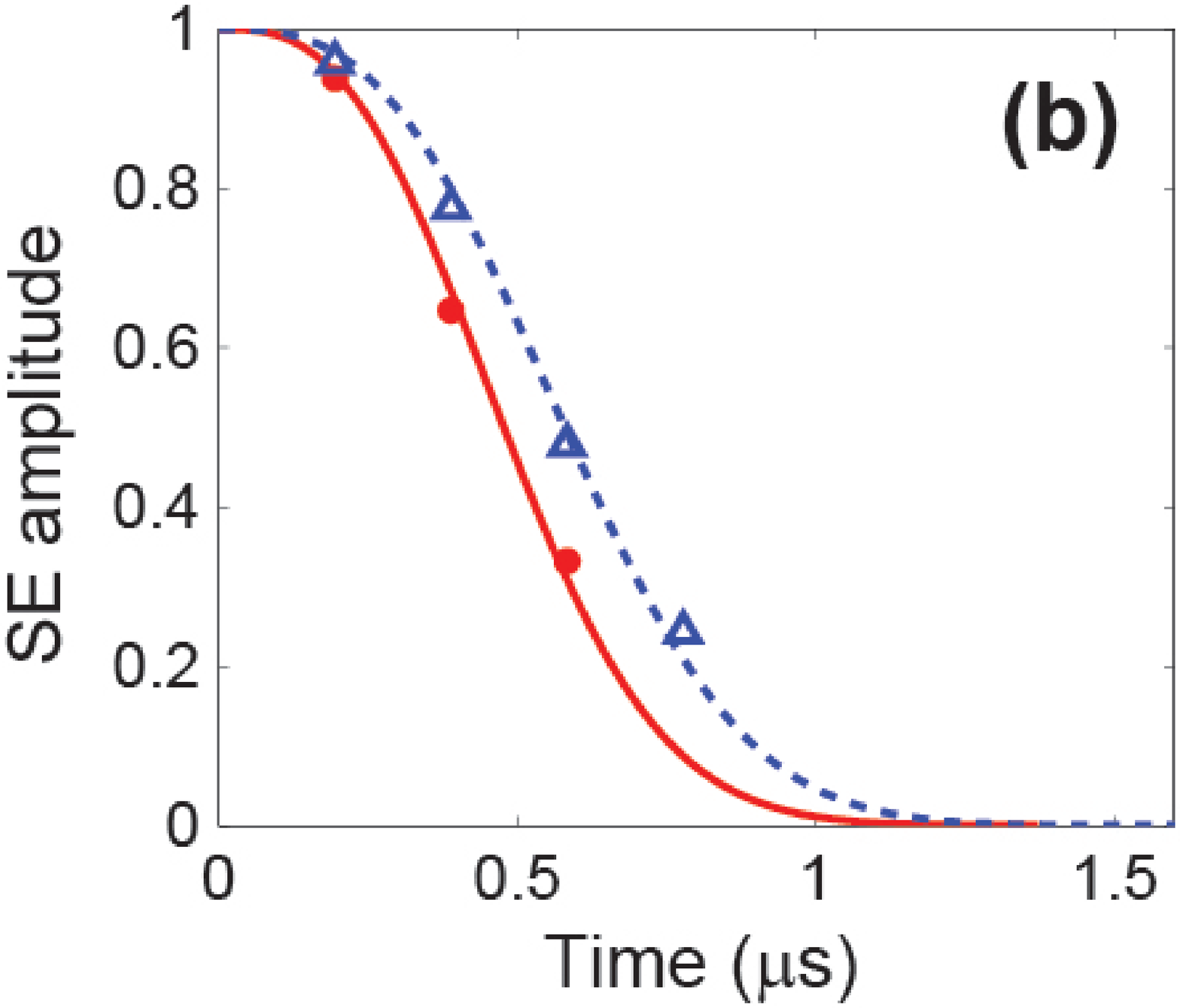}
\caption{
\label{fig:T2_groups}
Simulated SE signals with $f=50$~ppm of the N spin concentration.
(a) and (b) shows SE signals with two difference instances. Simulated results without (with) the HF coupling are shown by filled red circles (blue triangles).
Solid red (broken blue) lines are corresponding fits for the case without (with) the HF coupling.
}
\end{figure}
With the five energy levels for N spins with the magnetic field $B_0$ along the $[111]$ axis,
the number of N spins at the same energy level decreases, therefore the average distance between N spins with the same energy becomes larger.
This may result in a smaller $1/\tau_c$, and longer $T_2$ in SE signals.
Fig.~\ref{fig:T2_groups}(a) shows the simulated SE decays with and without the effect of the HF coupling with the same bath configuration (the number and location of N spins are same).
As shown in Fig.~\ref{fig:T2_groups}(a), the SE signals with and without the HF effect show no pronounced difference and both decays fit well with Eq.~(5) in the main text. On the other hand, in the other configuration, the SE signal with the HF coupling shows a longer $T_2$, as shown in Fig.~\ref{fig:T2_groups}(b).
We simulated 80 instances for statistical analysis of the HF coupling effect.
Fig.~3 in the main text shows the effect of the HF coupling with the distribution due to the spin bath configuration for $f=50~$ppm.
The obtained $1/T_2$ is 3.67 $\pm$ 2.55 $\mu s^{-1}$ and 3.35 $\pm$ 2.59 $\mu s^{-1}$ with and without the HF coupling.
Thus, we found that the effect of the HF is much smaller than the deviation due to the spin bath configuration.

{\bf (iii) Spin bath configuration dependence of the spin-bath coupling constant $b$}

The value of $b$ depends on the configuration of N spins in diamond. The distribution of the spin-bath coupling constant $b$ is given by~\cite{Dobrovitski08, Abragam},
\begin{eqnarray}
\label{eq:distr_b}
P(b)=\frac{\Gamma}{b^2}\sqrt\frac{2}{\pi}\exp[-\Gamma^2/(2b^2)]\;
\end{eqnarray}
where $\Gamma$ is determined by the decay rate of FID in an ensemble measurement.
\begin{figure}[htbp]
\includegraphics[width=10cm]{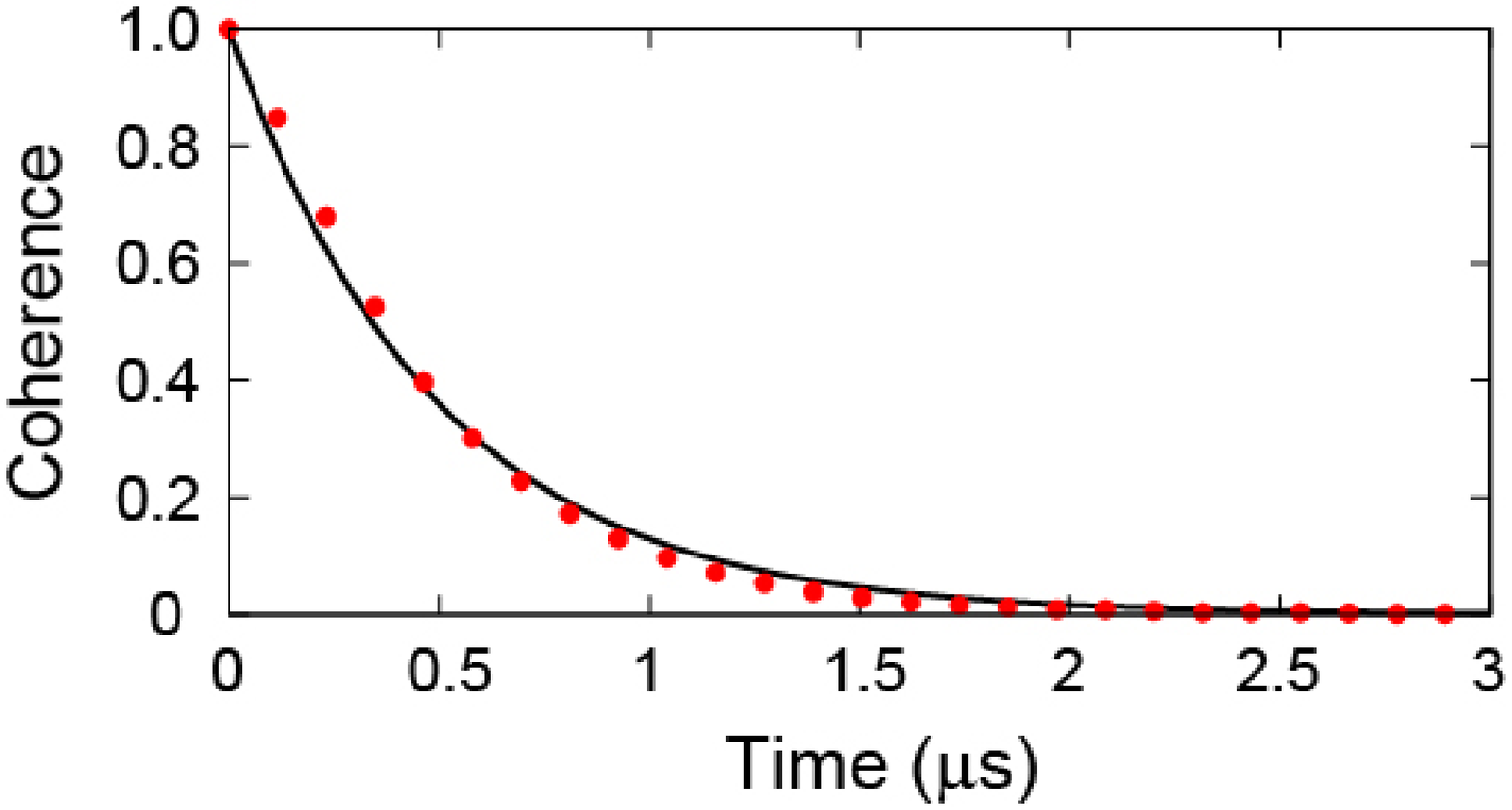}
\caption{\label{fig:T2Star_av}
Simulated FID for ensemble of NV spins with $f=10$~ppm of the N spin concentration (red dots). A black line represents a fit to a single exponential function giving $T_2^*=1/\Gamma=0.5~\mu s$.
}
\end{figure}
Fig.~\ref{fig:T2Star_av} shows a simulated ensemble FID result of the NV centers to determine $\Gamma$. The ensemble FID result is obtained by averaging over FID results of a single NV center from 90 difference configurations.